\documentclass[a4paper,11pt]{article}	

\usepackage{fancyhdr, graphicx}
\usepackage{float}
\usepackage{color}
\usepackage{amsmath,amssymb,wasysym}
\usepackage{multirow}
\usepackage{epstopdf}
\usepackage{lineno}
\usepackage[numbers,sort&compress]{natbib}
\usepackage{url}
\usepackage{authblk}
\usepackage[margin=1in]{geometry}

\title{First measurement of elastic, inelastic and total cross-section at $\sqrt{s}=13$~TeV by TOTEM and overview of cross-section data at LHC energies\vspace{5mm}}
\newif\ifFirstAuthor
\FirstAuthortrue

\def\AddAuthor#1#2#3#4{%
	\def\PriAf{#2}%
	\def\SecAf{#3}%
	\def\ExtAf{#4}%
	\def\empty{}%
	\ifx\PriAf\empty
		\author[]{#1$^{#4)}$}%
	\else
		\ifx\SecAf\empty
			\ifx\ExtAf\empty
				\author[]{#1$^{#2)}$}%
			\else
				\author[]{#1$^{#2,#4)}$}%
			\fi
		\else
			\ifx\ExtAf\empty
				\author[]{#1$^{#2,#3)}$}%
			\else
				\author[]{#1$^{#2,#3,#4)}$}%
				\relax
			\fi
		\fi
	\fi
}

\def\AddInstitute#1#2{%
	\affil[]{\scriptsize $^{#1}$#2\vspace{-3mm}}
}

\def\AddExternalInstitute#1#2{%
	{$^{\rm #1}$ #2}\\%
}

\def\DeclareAuthors{%
	\AddAuthor{G.~Antchev}{}{}{a}%
	\AddAuthor{P.~Aspell}{9}{}{}%
	\AddAuthor{I.~Atanassov}{}{}{a}%
	\AddAuthor{V.~Avati}{7}{9}{}%
	\AddAuthor{J.~Baechler}{9}{}{}%
	\AddAuthor{C.~Baldenegro~Barrera}{11}{}{}%
	\AddAuthor{V.~Berardi}{4a}{4b}{}%
	\AddAuthor{M.~Berretti}{2a}{}{}%
	\AddAuthor{E.~Bossini}{6b}{}{}%
	\AddAuthor{U.~Bottigli}{6b}{}{}%
	\AddAuthor{M.~Bozzo}{5a}{5b}{}%
	\AddAuthor{H.~Burkhardt}{9}{}{}%
	\AddAuthor{F.~S.~Cafagna}{4a}{}{}%
	\AddAuthor{M.~G.~Catanesi}{4a}{}{}%
	\AddAuthor{M.~Csan\'{a}d}{3a}{}{b}%
	\AddAuthor{T.~Cs\"{o}rg\H{o}}{3a}{3b}{}%
	\AddAuthor{M.~Deile}{9}{}{}%
	\AddAuthor{F.~De~Leonardis}{4c}{4a}{}%
	\AddAuthor{A.~D'Orazio}{4c}{4a}{}%
	\AddAuthor{M.~Doubek}{1c}{}{}%
	\AddAuthor{D.~Druzhkin}{9}{}{}%
	\AddAuthor{K.~Eggert}{10}{}{}%
	\AddAuthor{V.~Eremin}{}{}{e}%
	\AddAuthor{F.~Ferro}{5a}{}{}%
	\AddAuthor{A.~Fiergolski}{9}{}{}%
	\AddAuthor{F.~Garcia}{2a}{}{}%
	\AddAuthor{V.~Georgiev}{1a}{}{}%
	\AddAuthor{S.~Giani}{9}{}{}%
	\AddAuthor{L.~Grzanka}{7}{}{c}%
	\AddAuthor{J.~Hammerbauer}{1a}{}{}%
	\AddAuthor{J.~Heino}{2a}{}{}%
	\AddAuthor{P.~Helander}{2a}{2b}{}%
	\AddAuthor{T.~Isidori}{11}{}{}%
	\AddAuthor{V.~Ivanchenko}{8}{}{}%
	\AddAuthor{M.~Janda}{1c}{}{}%
	\AddAuthor{A.~Karev}{9}{}{}%
	\AddAuthor{J.~Ka\v{s}par}{6a}{1b}{}%
	\AddAuthor{J.~Kopal}{9}{}{}%
	\AddAuthor{V.~Kundr\'{a}t}{1b}{}{}%
	\AddAuthor{S.~Lami}{6a}{}{}%
	\AddAuthor{G.~Latino}{6b}{}{}%
	\AddAuthor{R.~Lauhakangas}{2a}{}{}%
	\AddAuthor{R.~Linhart}{1a}{}{}%
	\AddAuthor{C.~Lindsey}{11}{}{}%
	\AddAuthor{M.~V.~Lokaj\'{\i}\v{c}ek}{1b}{}{}%
	\AddAuthor{L.~Losurdo}{6b}{}{}%
	\AddAuthor{M.~Lo~Vetere}{5b}{5a}{+}%
	\AddAuthor{F.~Lucas~Rodr\'{i}guez}{9}{}{}%
	\AddAuthor{M.~Macr\'{\i}}{5a}{}{}%
	\AddAuthor{M.~Malawski}{7}{}{c}%
	\AddAuthor{N.~Minafra}{11}{}{}%
	\AddAuthor{S.~Minutoli}{5a}{}{}%
	\AddAuthor{T.~Naaranoja}{2a}{2b}{}%
	\AddAuthor{F.~Nemes}{9}{3a}{}%
	\AddAuthor{H.~Niewiadomski}{10}{}{}%
	\AddAuthor{T.~Nov\'{a}k}{3b}{}{}%
	\AddAuthor{E.~Oliveri}{9}{}{}%
	\AddAuthor{F.~Oljemark}{2a}{2b}{}%
	\AddAuthor{M.~Oriunno}{}{}{f}%
	\AddAuthor{K.~\"{O}sterberg}{2a}{2b}{}%
	\AddAuthor{P.~Palazzi}{9}{}{}%
	\AddAuthor{V.~Passaro}{4c}{4a}{}%
	\AddAuthor{Z.~Peroutka}{1a}{}{}%
	\AddAuthor{J.~Proch\'{a}zka}{1b}{}{}%
	\AddAuthor{M.~Quinto}{4a}{4b}{}%
	\AddAuthor{E.~Radermacher}{9}{}{}%
	\AddAuthor{E.~Radicioni}{4a}{}{}%
	\AddAuthor{F.~Ravotti}{9}{}{}%
	\AddAuthor{E.~Robutti}{5a}{}{}%
	\AddAuthor{C.~Royon}{11}{}{}%
	\AddAuthor{G.~Ruggiero}{9}{}{}%
	\AddAuthor{H.~Saarikko}{2a}{2b}{}%
	\AddAuthor{A.~Scribano}{6a}{}{}%
	\AddAuthor{J.~Siroky}{1a}{}{}%
	\AddAuthor{J.~Smajek}{9}{}{}%
	\AddAuthor{W.~Snoeys}{9}{}{}%
	\AddAuthor{R.~Stefanovitch}{9}{}{}%
	\AddAuthor{J.~Sziklai}{3a}{}{}%
	\AddAuthor{C.~Taylor}{10}{}{}%
	\AddAuthor{E.~Tcherniaev}{8}{}{}%
	\AddAuthor{N.~Turini}{6b}{}{}%
	\AddAuthor{V.~Vacek}{1c}{}{}%
	\AddAuthor{J.~Welti}{2a}{2b}{}%
	\AddAuthor{J.~Williams}{11}{}{}%
	\AddAuthor{P.~Wyszkowski}{7}{}{}%
	\AddAuthor{J.~Zich}{1a}{}{}%
	\AddAuthor{K.~Zielinski}{7}{}{}%
}

\def\DeclareInstitutes{%
	\AddInstitute{1a}{University of West Bohemia, Pilsen, Czech Republic.}
	\AddInstitute{1b}{Institute of Physics of the Academy of Sciences of the Czech Republic, Prague, Czech Republic.}
	\AddInstitute{1c}{Czech Technical University, Prague, Czech Republic.}
	\AddInstitute{2a}{Helsinki Institute of Physics, University of Helsinki, Helsinki, Finland.}
	\AddInstitute{2b}{Department of Physics, University of Helsinki, Helsinki, Finland.}
	\AddInstitute{3a}{Wigner Research Centre for Physics, RMKI, Budapest, Hungary.}
	\AddInstitute{3b}{EKU KRC, Gy\"ongy\"os, Hungary.}
	\AddInstitute{4a}{INFN Sezione di Bari, Bari, Italy.}
	\AddInstitute{4b}{Dipartimento Interateneo di Fisica di Bari, Bari, Italy.}
	\AddInstitute{4c}{Dipartimento di Ingegneria Elettrica e dell'Informazione - Politecnico di Bari, Bari, Italy.}
	\AddInstitute{5a}{INFN Sezione di Genova, Genova, Italy.}
	\AddInstitute{5b}{Universit\`{a} degli Studi di Genova, Italy.}
	\AddInstitute{6a}{INFN Sezione di Pisa, Pisa, Italy.}
	\AddInstitute{6b}{Universit\`{a} degli Studi di Siena and Gruppo Collegato INFN di Siena, Siena, Italy.}
	\AddInstitute{7}{AGH University of Science and Technology, Krakow, Poland.}
	\AddInstitute{8}{Tomsk State University, Tomsk, Russia.}
	\AddInstitute{9}{CERN, Geneva, Switzerland.}
	\AddInstitute{10}{Case Western Reserve University, Dept.~of Physics, Cleveland, OH, USA.}
	\AddInstitute{11}{The University of Kansas, Lawrence, USA.}
}

\def\DeclareExternalInstitutes{%
	\AddExternalInstitute{a}{INRNE-BAS, Institute for Nuclear Research and Nuclear Energy, Bulgarian Academy of Sciences, Sofia, Bulgaria.}
	\AddExternalInstitute{b}{Department of Atomic Physics, ELTE University, Budapest, Hungary.}
	\AddExternalInstitute{c}{Institute of Nuclear Physics, Polish Academy of Science, Krakow, Poland.}
	\AddExternalInstitute{d}{Warsaw University of Technology, Warsaw, Poland.}
	\AddExternalInstitute{e}{Ioffe Physical - Technical Institute of Russian Academy of Sciences, St.~Petersburg, Russian Federation.}
	\AddExternalInstitute{f}{SLAC National Accelerator Laboratory, Stanford CA, USA.}
	\AddExternalInstitute{+}{Deceased.}
}

\DeclareAuthors
\DeclareInstitutes
\begin{document}
%\date{}
\maketitle
\newpage
%\title{First measurement of elastic, inelastic, total cross-section at $\sqrt{s}=13$~TeV by TOTEM and overview of cross-section data at LHC energies}
%\author{TOTEM collaboration (G.~Antchev \emph{\etal})}

\begin{abstract}
The TOTEM collaboration has measured the proton-proton total cross section at $\sqrt{s}=13$~TeV  with a luminosity-independent method. Using dedicated $\beta^{*}=90$~m beam optics, the Roman
Pots were inserted very close to the beam. The inelastic scattering rate has been measured by the T1 and T2 telescopes during the same LHC fill.
After applying the optical theorem the total proton-proton cross section is $\sigma_{\rm tot}=(110.6~\pm~3.4$)~mb, well in agreement with the extrapolation from lower energies. This method also allows one to derive the luminosity-independent elastic and inelastic cross sections:
$\sigma_{\rm el}=(31.0~\pm~1.7)$~mb and $\sigma_{\rm inel}=(79.5~\pm~1.8)$~mb.
\end{abstract}

\vspace{160mm}
\begin{flushleft}\scriptsize
\bigskip
\DeclareExternalInstitutes\bigskip
\end{flushleft}
\vfill\eject

%\end{titlepage}

\section{Introduction}

This paper presents the first measurement of the total proton-proton cross section at a center of mass energy $\sqrt{s}=13$~TeV; the measurement is luminosity independent. 

The TOTEM collaboration has already measured the total
proton-proton cross section at $\sqrt{s}=2.76$~TeV, 7~TeV and 8~TeV, and has demonstrated the reliability of the luminosity-independent
method by comparing several approaches to determine
the total cross sections~\cite{Antchev:2013paa,Antchev:2016vpy,Antchev:2011vs,Antchev:2013iaa,DIS2017_proceedings,Paper_2p76}. The method requires the
simultaneous measurements of the inelastic and elastic
rates, as well as the extrapolation of the latter in the
invisible region down to vanishing four-momentum transfer squared $t=0$. 

%This is
%achieved with the TOTEM experimental setup which
%consists of two inelastic telescopes T1 and T2 to detect
%charged particles produced in inelastic $pp$ collisions, and
%Roman Pot stations to detect elastically scattered protons
%at very small angles.

The TOTEM experimental setup consists of two inelastic telescopes T1 and T2 to detect charged particles coming from inelastic $\rm pp$ collisions and
the Roman Pot detectors (RP) to detect elastically scattered protons at very small angles.
The inelastic telescopes are placed symmetrically on both sides of Interaction Point 5 (IP5): the T1
telescope is based on cathode strip chambers (CSCs)
placed at $\pm$9~m and covers the pseudorapidity range 3.1~$\le |\eta| \le$~4.7; the T2 telescope is based on gas electron
multiplier (GEM) chambers placed at $\pm$13.5~m and covers the pseudorapidity range 5.3~$\le |\eta| \le$~6.5. The pseudorapidity coverage of the two telescopes at $\sqrt{s}=13$~TeV allows the detection of about 92~\% of the inelastic events, including events with diffractive mass
down to 4.6 GeV. As the fraction of events with all final state particles beyond the instrumented region
has to be estimated using phenomenological models, the
excellent acceptance in TOTEM minimizes the dependence on such models and thus provides small uncertainty on the
inelastic rate measurement.

			The Roman Pot (RP) units used for the present measurement are located on both sides of the IP at distances of $\pm213$~m (near) and $\pm220$~m (far) from IP5. A unit consists of 3 RPs, two approaching the outgoing beam vertically and one horizontally.
			The horizontal RP overlaps with the two verticals and allows for a precise relative alignment of
			the detectors within the unit. The $7$~m long lever arm between the near and the far RP units has the important advantage that the local track angles in the $x$ and $y$-projections perpendicular to the beam direction can be reconstructed with a precision of 2~$\mu$rad. A complete description of the TOTEM detector is given in~\cite{Anelli:2008zza,TOTEM:2013iga}.
			
			Each RP is equipped with a stack of 10 silicon strip detectors designed with the specific objective of reducing the insensitive area at the edge facing the beam to only a few tens of micrometers. The 512 strips with 66
			$\mu$m pitch of each detector are oriented at an angle of +45$^{\circ}$ (five planes) and -45$^{\circ}$ (five planes) with respect to the
			detector edge facing the beam~\cite{Ruggiero:2009zz}.

\section{Data taking and analysis}
{\color{black} The analysis is performed on two data samples (DS1 and DS2) recorded in 2015 during a special LHC fill with $\beta^{*}=90$~m optics.
This special optics configuration is described in detail in~\cite{Antchev:2011vs,Antchev:2013gaa,Nemes:2131667,Antchev:2014voa}.

The RP detectors were placed as close as 5 times the transverse beam size ($\sigma_{\rm beam}$) from the outgoing beams. The collected events have been triggered by the T2
telescope in either arm (inelastic trigger), by the RP detectors in a double-arm coincidence (elastic trigger), and by
random bunch crossings (zero-bias sample used for calibration). In DS2 there are no zero-bias data recorded, and the closest run with zero-bias
data is used for calibration; the time dependence of the zero-bias trigger rate is taken into account with a scale factor measured on the physics data of DS2 itself and the closest run.
}

	\subsection{Elastic analysis}

	\subsubsection{Reconstruction of kinematics}
		The horizontal and vertical scattering angles of the proton at IP5 $(\theta_{x}^{*},\theta_{y}^{*})$  are reconstructed in a given arm by inverting the proton transport
		equations~\cite{Antchev:2014voa}
			\begin{align}
			    \theta_{x}^{*} = \frac{1}{\frac{{\rm d}L_{x}}{{\rm d}s}}\left(\theta_{x}-\frac{{\rm d} v_{x}}{{\rm d} s}x^{*}\right)\,,\,
    			    \theta^{*}_{y} = \frac{y}{L_{y}}\,,
			    \label{reconstruction_formula_theta_x_rearranged}
			\end{align}
			where $s$ denotes the distance from the interaction point, $y$ is the vertical coordinate of the proton's trajectory, $\theta_{x}$ is its horizontal
			angle at the detector, and $x^{*}$ is the horizontal vertex coordinate reconstructed as
			\begin{align}
			    x^{*}&=\frac{L_{x,{\rm far}}\cdot x_{\rm near} - L_{x,{\rm near}}\cdot x_{\rm far}}{d}\,,
			    \label{reconstruction_formula_x}
			\end{align}
			where $d=( v_{x,{\rm near}}\cdot L_{x,{\rm far}} -  v_{x,{\rm far}}\cdot L_{x,{\rm near}})$. The scattering angles obtained for the two arms are averaged 
			and the four-momentum transfer squared is calculated
			\begin{align}
			    t=-p^{2}\theta^{*2}\,,
			    \label{reconstructed_t}
			\end{align}
		where $p$ is the LHC beam momentum and the scattering angle $\theta^{*}=\sqrt{{\theta_{x}^{*}}^{2} + {\theta_{y}^{*}}^{2}}$.

		The coefficients $L_{x}$, $L_{y}$ and $v_{x}$ of Eq.~(\ref{reconstruction_formula_theta_x_rearranged}) and Eq.~(\ref{reconstruction_formula_x}) are optical functions of the LHC beam determined by the
		accelerator magnets. The $\beta^{*}=90$~m optics is designed with a large vertical effective length $L_{y}\approx~263$~m at the RPs placed at $220$~m from IP5. Since the horizontal
		effective length $L_{x}$ is close to zero at the RPs, its derivative ${{\rm d}L_{x}}/{{\rm d}s}\approx-0.6$ is used instead. The different reconstruction formula in the vertical
		and horizontal plane in Eq.~(\ref{reconstruction_formula_theta_x_rearranged}) is also motivated by their different sensitivity to LHC magnet and beam perturbations.

	\subsubsection{RP alignment and beam optics}
	\label{RP_alignment}

	After applying the usual TOTEM alignment methods the residual misalignment is about 10~$\mu$m in the horizontal coordinate and about 150~$\mu$m in the vertical~\cite{Antchev:2016vpy,Antchev:2015zza}. When propagated to the
	reconstructed scattering angles, this leads to uncertainties of about 3.4~$\mu$rad (horizontal angle) and 0.6~$\mu$rad (vertical angle). The beam
	divergence uncertainty has been convoluted with the vertical alignment for the error propagation.

	The nominal optics has been updated from LHC magnet and current databases and
	calibrated using the observed elastic candidates. The uncertainties of the optical functions are estimated with a Monte Carlo program applying the optics calibration procedure
	on a sophisticated simulation of the LHC beam and its perturbations. The obtained uncertainty is about 1.2~$\permil$ for ${{\rm d}L_{x}}/{{\rm d}s}$ and $2.1~\permil$ for $L_{y}$~\cite{Antchev:2014voa,Nemes:2131667}.

	The statistical uncertainty of the scattering angles, obtained from the data, is $1.9\pm0.1~\mu$rad vertically (mainly due to the beam divergence) and $4.9\pm0.1~\mu$rad horizontally (due to the beam divergence and sensor pitch).

	\subsubsection{Event selection}

	\begin{figure}
		\centering
		\includegraphics[width=0.75\columnwidth]{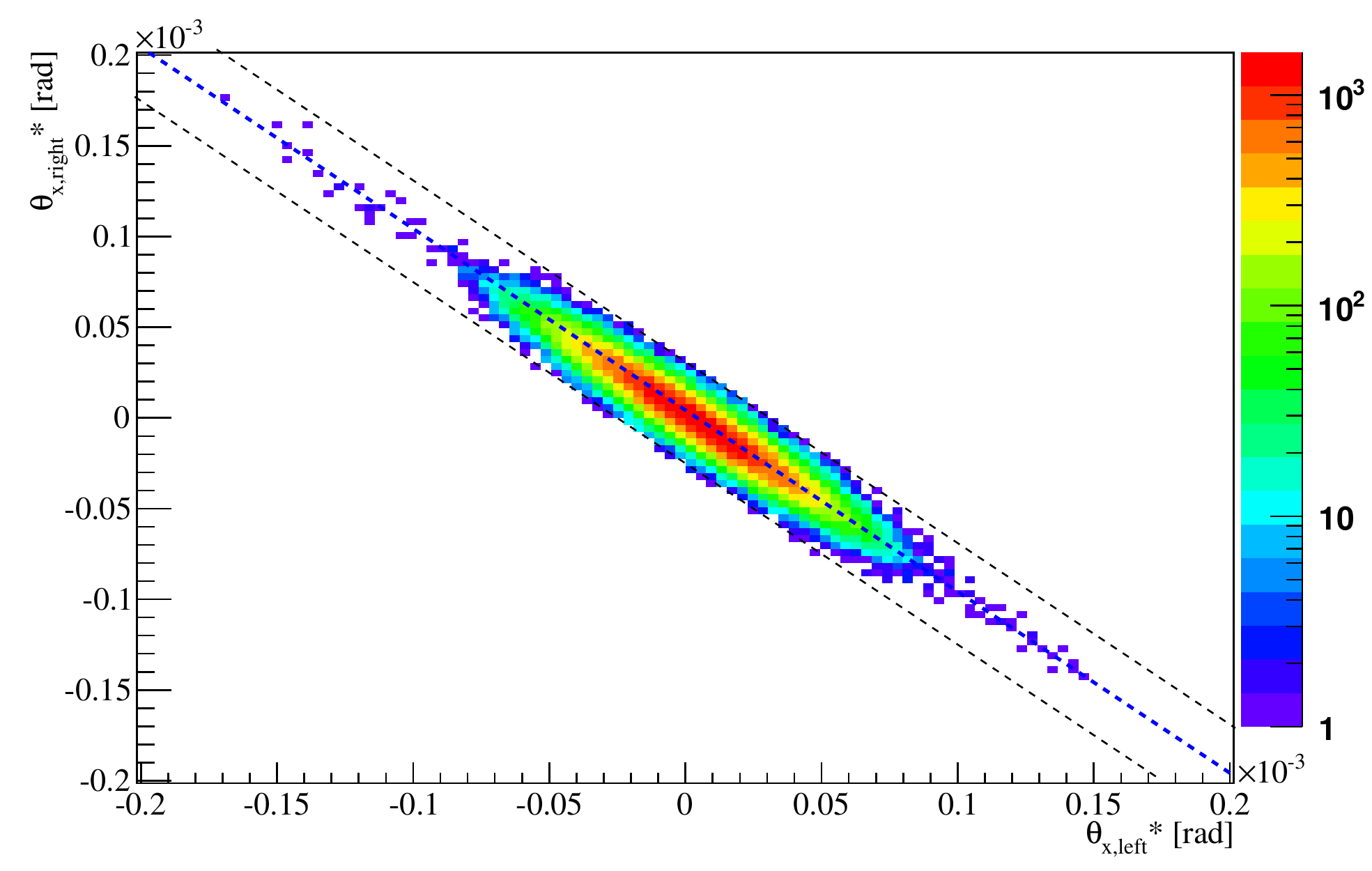}
		\caption{(color) Analysis cut in the horizontal scattering angle $\theta_{x}^{*}$. The blue and black dashed lines represent the mean and the 4$\sigma$ cuts, respectively.}
		\label{cuts}
 	\end{figure}

	The analysis is similar to the procedure performed for the measurement of the elastic cross section at several other LHC energies: 2.76~TeV, 7~TeV and 8~TeV~\cite{Antchev:2013paa,Antchev:2016vpy,Antchev:2011vs,Antchev:2013iaa,DIS2017_proceedings,Paper_2p76}. 
	The measurement of the elastic rate is based on the selection of events with the following topology in the RP detector system: a reconstructed track in the near and far
	vertical detectors on each side of the IP such that the elastic signature is satisfied in one of the two diagonals: left bottom and right top (Diag. 1) or
	left top and right bottom (Diag. 2).

	Besides, the elastic event selection requires the collinearity of the outgoing protons in the two arms, the suppression of the
	diffractive events and the equality of the horizontal vertex position $x^{*}$ reconstructed from the left and right
	arms.

	\begin{figure}[H]
		\centering
		\includegraphics[width=0.75\columnwidth]{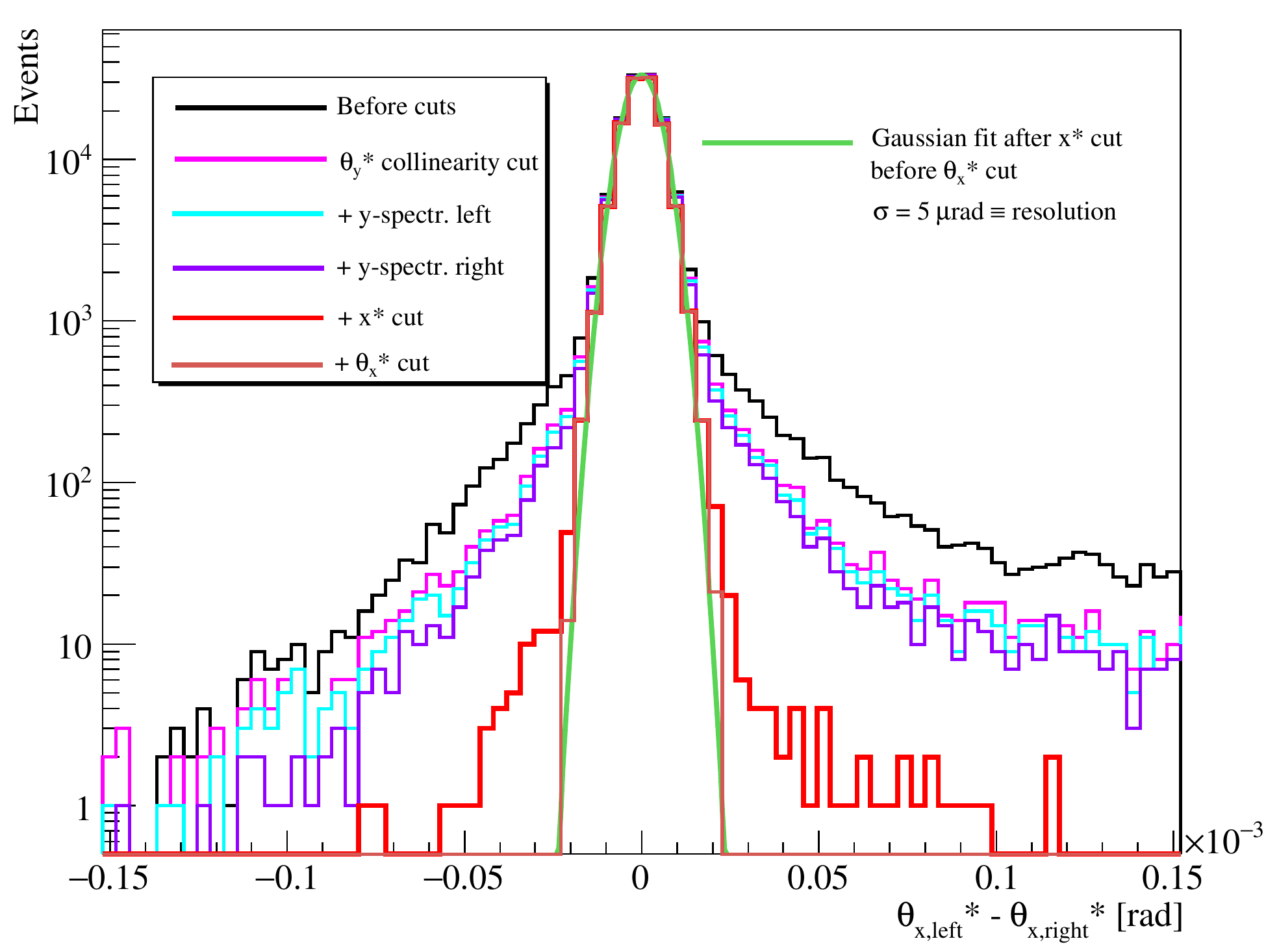}
		\caption{(color) The distribution of the horizontal scattering angle difference reconstructed from the left and the right arm. The distribution is shown before
		any analysis cut (black solid line) and after each analysis cut.}
		\label{signal_to_noise}
	\end{figure}

	Figure~\ref{cuts} shows the horizontal collinearity cut imposing momentum conservation in the horizontal plane with 1~$\permil$ uncertainty. The cuts are applied at the 4$\sigma$ level, and they are optimized for
	purity (background contamination in the selected sample	less than 0.1~\%) and for efficiency (uncertainty of true elastic event selection 0.5~\%). Figure~\ref{signal_to_noise} shows the progressive selection
	of elastic events after each analysis cut. 

	\subsubsection{Geometrical and beam divergence correction, unfolding}
		{\color{black} The acceptance of elastically scattered protons is limited by the RP silicon detector edge and by the LHC magnet apertures. The proton acceptance correction is calculated taking into account
		the azimuthal symmetry of elastic scattering, experimentally verified on the data 
		\begin{equation}
			\mathcal{A}(\theta^{*})=\frac{2\pi }{\Delta \phi^{*}(\theta^{*})}\,,
		\end{equation}
 		where $\Delta \phi^{*}$ is the visible azimuthal angle range, defined by the acceptance cuts. The $t$-range of the analysis is constrained to
		$|t|_{\rm min}=1.2\cdot10^{-2}$~GeV$^{2}$ and $|t|_{\rm max}=0.2$~GeV$^{2}$, a region where the acceptance correction factor
		$\mathcal{A}(\theta^{*})$ is below six in order to limit the systematic error on the final cross section.

 		Close to the acceptance edges the assumed azimuthal symmetry has to be corrected due to the beam divergence. This additional acceptance loss is modelled with a Gaussian distribution,
		with experimentally determined parameters, and taken into account as a function of the vertical scattering angle $\mathcal{D}(\theta_{y}^{*})$.

		The unfolding of resolution effects is estimated with a Monte Carlo simulation whose parameters are obtained
		from the data, see Section~\ref{RP_alignment}. The angular spread of the beam is determined with an
		uncertainty 0.1 $\mu$rad by comparing the scattering angles reconstructed from the left and right arm, therefore 
    		the unfolding correction factor $\mathcal{U}(\theta^{*})$ can be calculated with a precision better than 0.1~\%.}
 		The event-by-event correction factor due to acceptance corrections and resolution unfolding is
		\begin{equation}
			\mathcal{C}(\theta^{*},\theta_{y}^{*})=\mathcal{A}(\theta^{*})\mathcal{D}(\theta_{y}^{*})\mathcal{U}(\theta^{*})\,,
		\end{equation}
		see Table~\ref{analysis_corrections}.
 
	\subsubsection{Inefficiency corrections}
		{\color{black} The proton reconstruction efficiency of the RP detectors is evaluated directly from the data. The strip detectors
		are not able to resolve multiple tracks, which is the main source of detector inefficiency. The additional tracks can be caused by interactions of
		the protons with the sensors or the surrounding  material or by the pileup with non-signal protons.

		The inefficiencies corrections are calculated for different categories: ``uncorrelated'' ($\mathcal{I}_{\rm 3/4}$) when one RP out of four along a diagonal
		has no reconstructed track; this inefficiency includes the loss of the track due to nuclear interaction, shower or pile-up with beam halo and is calculated as a function of 
		$\theta_{y}^{*}$ per RP~\cite{Antchev:2015zza}. The inefficiency is called
		``correlated'' ($\mathcal{I}_{\rm 2/4}$) when both RP of one arm have no reconstructed tracks. The case when two RPs have no reconstructed track in two different
		arms ($\mathcal{I}_{\rm 2/4\,diff.}$) is derived with a probability formula from the ``uncorrelated'' inefficiency.
		The numerical values of these corrections are listed in Table~\ref{analysis_corrections_summary}.

		The total correction factor per event is
		\begin{equation}
			f(\theta^{*},\theta_{y}^{*})=\frac{1}{\eta_{\rm d}\eta_{\rm tr}}\cdot\frac{\mathcal{C}(\theta^{*},\theta_{y}^{*})}{1-\mathcal{I}}\cdot\frac{1}{\Delta t}\,,
		\end{equation}
		where the track reconstruction inefficiencies are summed $\mathcal{I}=\mathcal{I}_{3/4}(\theta_{y}^{*}) + \mathcal{I}_{\rm 2/4}+\mathcal{I}_{\rm 2/4\,diff}$ since they are
		mutually exclusive, $\Delta t$ is the bin width and $\eta_{\rm d}$, $\eta_{\rm tr}$ are the DAQ and trigger efficiency, respectively. The observed $N_{\rm el,obs}$ and the fully corrected elastic rate $N_{\rm el}$ is
		summarized for the two data sets in Table~\ref{data_reduction}, together with their optical point $\left.{\rm d}N_{\rm el}/{\rm d}t\right|_{t=0}$ 
		.

%     \begin{table}
%         \begin{center}
%             \begin{tabular}{ | c | c | }
% 		\hline
%                 	Method			& Number of events 	\\ \hline\hline
% 			Total RP triggers	& 	\\ \hline
% 			Reconstructed tracks	& 	\\ \hline
% 			Elastic events		& 	\\ \hline
%             \end{tabular}
%             \caption{Data reduction with cuts.}
%         \label{data_reduction}
%         \end{center}
%     \end{table}

    \begin{table*}\small\color{black}
        \begin{center}
            \caption{The $t$-dependent analysis uncertainties and corrections to the differential elastic rate and to the optical point for both diagonals.}
            \begin{tabular}{ | c  c  c |}
		\hline
                					 				& $|t|_{\rm min}$				& $|t|_{\rm max}$					\\ \hline 
			Alignment uncertainty						& $\pm1.3~\%$					& $\pm3.4~$\%						\\ 
			Optics uncertainty						& $\pm1\permil$					& $\pm1\permil$						\\ 
			$\mathcal{A}(\theta^{*})$                 			& $5.96\pm3\times10^{-2}$ 			& $1.03\pm1\times10^{-2}$	                           		\\  
			$\mathcal{D}(\theta_{y}^{*})$					& $2.31\pm2\times10^{-2}$			& $1.00\pm1\times10^{-3}$						\\ 
			$\mathcal{U}(\theta^{*})$					& $1.002\pm5\times10^{-5}$	 		& $1.04\pm8\times10^{-4}$ 		\\ \hline
            \end{tabular}
        \label{analysis_corrections}
        \end{center}
    \end{table*}

    \begin{table*}\small\color{black}
        \begin{center}
            \caption{Corrections to the differential and total elastic rate for the different datasets and diagonals. The ``uncorrelated'' inefficiency correction ($\mathcal{I}_{\rm 3/4}$)
		is $\theta_{y}^{*}$ dependent, in the table its effect on the elastic rate is provided.}
            \begin{tabular}{ | c  c  c  c  c |}
		\hline
				Correction~[\%]			 	 & \multicolumn{2}{c}{DS1}			& \multicolumn{2}{c|}{DS2}				\\ 
							 		 & Diag. 1		& Diag. 2		& Diag. 1		& Diag. 2	 			\\\hline\hline 
			$\mathcal{I}_{\rm 3/4}$		 		 & 25.86$~\pm~$0.2	& 22.04$~\pm~$0.2	& 20.34$~\pm~$0.1	& 21.37$~\pm~$0.1			\\ 
			$\mathcal{I}_{\rm 2/4}$		 		 & 19.91$~\pm~$0.2	& 16.16$~\pm~$0.2 	& 16.09$~\pm~$0.2	& 17.11$~\pm~$0.2			\\ 
			$\mathcal{I}_{\rm 2/4\,diff.}$   		 & 2.38$~\pm~$0.05	& 1.61$~\pm~$0.04	& 1.33$~\pm~$0.02	& 1.5$~\pm~$0.02  			\\ 
			$\eta_{\rm\scriptsize d}$	 		 & \multicolumn{2}{c}{80.93$~\pm~$0.01}      	& \multicolumn{2}{c|}{99.95$~\pm~$0.01}			\\ 
			$\eta_{\rm\scriptsize tr}$	 		 & \multicolumn{2}{c}{99.9$~\pm~$0.1}        	& \multicolumn{2}{c|}{99.9$~\pm~$0.1}				\\ \hline
            \end{tabular}
        \label{analysis_corrections_summary}
        \end{center}
    \end{table*}

\subsection{Analysis of inelastic scattering}

The analysis procedure is similar to the ones for the inelastic event rate measurements at 2.76, 7 and 8 TeV~\cite{DIS2017_proceedings,Antchev:2013haa,Antchev:2013paa,Paper_2p76} and starts from the
number of T2 triggered events as the observed inelastic rate. The events are classified according to their topology: events with tracks in T2 in both hemispheres (``2h''),
dominated by non-diffractive minimum bias and double diffraction, and events with tracks in one hemisphere only (``1h''), dominated by single diffraction. Due to the non-operational
 half-arm of T2 on the negative side, for the 1h category each of the three half-arms are treated separately in the analysis to avoid biases and then the two half-arms
 on the positive side are combined. 

To evaluate the total inelastic rate, several corrections have to be applied. First, to obtain the T2 visible inelastic rate (N$_{\rm T2vis}$), the observed rate is corrected for beam gas
 background, trigger, reconstruction efficiency and the effect of pileup. Next, the rate corresponding to the events with at least one final state particle in $|\eta|<6.5$ (N$_{|\eta|< 6.5}$) is derived
by assessing topologies which can cause an undetected event in T2. These are events detected only by T1, central diffractive events with all final state particles outside the T2 acceptance and events with
a local rapidity gap covering T2. Finally, to estimate the total inelastic rate (N$_{\rm inel}$), the contribution of low mass diffraction with only final state particles at $|\eta|>6.5$ is evaluated.
The corrections leading to the total inelastic rate measurement are described below and quantified in Table~\ref{inelastic_rate_corrections} together with their systematic uncertainties, summing up
to 3.7~\%. The observed $N_{\rm inel,obs}$ and fully corrected inelastic rate $N_{\rm inel}$ is shown in Table~\ref{data_reduction}.

	\begin{table*}
		\centering
            \caption{Corrections and systematic uncertainties of the inelastic rate measurement. The second column shows the size of the correction, the third column the systematic
		uncertainty related to the source.}
            \begin{tabular}{| c  c  c  c |}
		\hline\hline
                		Source				& 	Correction 	&	Uncertainty	& Effect on	\\ \hline
				Beam gas                        & -0.4~\%            	& 0.2~\%             	& all rates	\\ 
				Trigger efficiency              & 1.2~\%             	& 0.6~\%             	& all rates	\\ 
				Pile up                       	& 3.6~\%             	& 0.4~\%             	& all rates\\ 
				T2 event reconstruction    	& 0.9 (1.6)~\%     	& 0.45 (0.8)~\%   	& N$_{\rm inel}$, N$_{|\eta| < 6.5}$ (N$_{\rm T2vis}$) \\ 
				T1 only                         & 1.7~\%             	& 0.4~\%             	& N$_{\rm inel}$, N$_{|\eta| < 6.5}$ \\ 
				Central diffraction             & 0.5~\%             	& 0.35~\%           	& N$_{\rm inel}$, N$_{|\eta| < 6.5}$ \\ 
				Local rapidity gap covering T2 	& 0~\%               	& 0.4~\%             	& N$_{\rm inel}$, N$_{|\eta| < 6.5}$ \\ 
				Low mass diffraction seen       & -0.6~\%            	& 0.3~\%             	& N$_{\rm inel}$, N$_{|\eta| < 6.5}$ \\ 
				Low mass diffraction         	& 7.1~\%             	& 3.55~\%           	& N$_{\rm inel}$ \\ \hline
            \end{tabular}
	\label{inelastic_rate_corrections}
	\end{table*}

\subsubsection{Beam gas background}
The beam gas background is estimated from events triggered with T2 on the non-colliding bunches and affects only the 1h category. The intensity difference between the colliding and non-colliding bunches is taken
into account. Conservatively, half the size of the correction to the overall inelastic rate is taken as systematic uncertainty. 

\subsubsection{Trigger efficiency}
The trigger efficiency is determined from zero bias triggered events, separately for the different event topologies and integrated over all T2 track multiplicities. The systematic uncertainty is evaluated as
the variation required on the 1h trigger efficiency to give compatible fractions for left and right arm (after correcting for the non-operational half arm of T2).

\subsubsection{Pileup}
The pileup correction factor is determined from the zero bias triggered events. The probability to have a bunch crossing with tracks in T2 is about 0.07 from which the probability of having more than two inelastic collisions with tracks in T2 in the same bunch crossing is derived.
The systematic uncertainty is assessed from the variation of the probability, within the same dataset, to have a bunch crossing with tracks in T2 and the uncertainty due to the T2 event reconstruction efficiency.

\subsubsection{T2 event reconstruction}
The T2 event reconstruction inefficiency is estimated using Monte Carlo (MC) generators (PYTHIA8-4C~\cite{Corke:2010yf}, QGSJET-II-04~\cite{Ostapchenko:2013pia}) tuned with data to reproduce the measured fraction of 1h events, 0.195 $\pm$ 0.010. The systematic uncertainty is taken to be half of the correction that in these runs are mostly due to events with tracks only in the non-operational T2 half-arm with some additional events due to only neutral particles within the T2 acceptance. A large fraction of the events missed due to T2 reconstruction inefficiency are recuperated with the T1 detector reducing the correction sizably for N$_{\rm inel}$ and N$_{|\eta| < 6.5}$.

\subsubsection{T1 only}
The T1-only correction takes into account events with no reconstructed particles in T2 but tracks reconstructed in T1. The systematic uncertainty is equal to the precision to which this correction can be calculated from the zero-bias sample.

\subsubsection{Central diffraction}
The central diffraction correction, based on the PHOJET and MBR event generators~\cite{Engel:1996aa,Ciesielski:2012mc}, takes into account events with all final state particles outside the T1/T2 pseudorapidity acceptance. 
Both generators are underestimating the low mass resonance contribution. Therefore, the total central diffractive contribution is assumed to be twice the generator estimates. Since the uncertainties of the central
diffractive cross-section and the low mass resonance contribution are large, the systematic uncertainty is assumed to be equal to the largest difference of the correction with and without low mass resonance contribution.

\subsubsection{Local rapidity gap covering T2}
The correction due to local rapidity gap over T2 considers single diffraction events with a rapidity gap of the diffractive system extending over the entire T2 $\eta$-range and with no tracks in T1.
It is estimated from data, measuring the probability of having a single diffractive-like topology with a gap covering T1 and transferring it to the T2 region correcting for the different
conditions (average charged multiplicity, $p_T$ threshold, gap size and surrounding material) between T1 and T2. As a cross-check the correction is also estimated from MC generators
(PYTHIA8-4C, QGSJETII-04). The two estimates differ sizably and therefore only a systematic uncertainty equal to the largest estimate is applied, without making any correction.

\subsubsection{Low mass diffraction}
The T2 acceptance edge at $|\eta|$ = 6.5 corresponds to a diffractive mass of about 4.6 GeV (at 50~\% efficiency). The low mass diffraction correction, i.e. the contribution of events
with all final state particles at $|\eta|>6.5$, is estimated with QGSJET-II-03~\cite{Ostapchenko:2004ss} after correcting the fraction of 1h events in the MC to the one of the data. At 7 TeV, the estimated correction using this procedure was consistent
with the value estimated from data~\cite{Antchev:2013haa}. To account for the large uncertainty of the low mass diffraction contribution and to cover also other predictions~\cite{Corke:2010yf,Khoze:2009nc},
the systematic uncertainty is taken to be half of this correction.

\section{Cross sections}
		\subsection{Differential elastic rate and extrapolation to $t=0$}

			After unfolding and including all systematic uncertainties, the differential elastic rate ${\rm d}N_{\rm el}/{\rm d}t$ is described with an exponential
			and fitted in the $|t|_{\rm min}$ and $|t|_{\rm max}$ range, see Figure~\ref{differential_cross_section}. The normalized $\chi^{2}/{\rm ndf}=50.8/36=1.4$ is representative
			of the known deviations from a pure exponential~\cite{Antchev:2015zza}.

	\begin{table*}
		\centering
        	\caption{The observed elastic $N_{\rm el,obs}$ and inelastic rate $N_{\rm inel,obs}$, the fully corrected elastic $N_{\rm el}$ and inelastic rate $N_{\rm inel}$
		and the optical point $\left.{\rm d}N_{\rm el}/{\rm d}t\right|_{t=0}$ of the two data sets (errors where quoted are statistical and systematic).}
	        \begin{tabular}{| c c  c  c |}
		\hline
               		Data set				& Unit		& DS1 					& DS2						\\ \hline\hline
			$N_{\rm el,obs}$			&		& 105729				& 216825					\\ 
			$N_{\rm inel,obs}$			&		& 773000				& 1488343					\\  
			$N_{\rm el}$				&		& $4.273\cdot10^{5}\pm0.5~\%\pm2.3$~\%	& $6.660\cdot10^{5}\pm0.5~\%\pm2.3$~\%		\\ 
			$\left.{\rm d}N_{\rm el}/{\rm d}t\right|_{t=0}$ 	& [GeV$^{-2}$]	& $8.674\cdot10^{6}\pm0.4~\%\pm1.6$~\%	& $1.356\cdot10^{7}\pm0.4~\%\pm1.6$~\%		\\ 
			$N_{\rm inel}$				&		& $1.097\cdot10^{6}\pm0.1~\%\pm3.7~\%$	& $1.708\cdot10^{6}\pm0.1~\%\pm3.7~\%$		\\ \hline 
        	\end{tabular}
		\label{data_reduction}
	\end{table*}

	\begin{figure*}
		\includegraphics[width=0.515\linewidth]{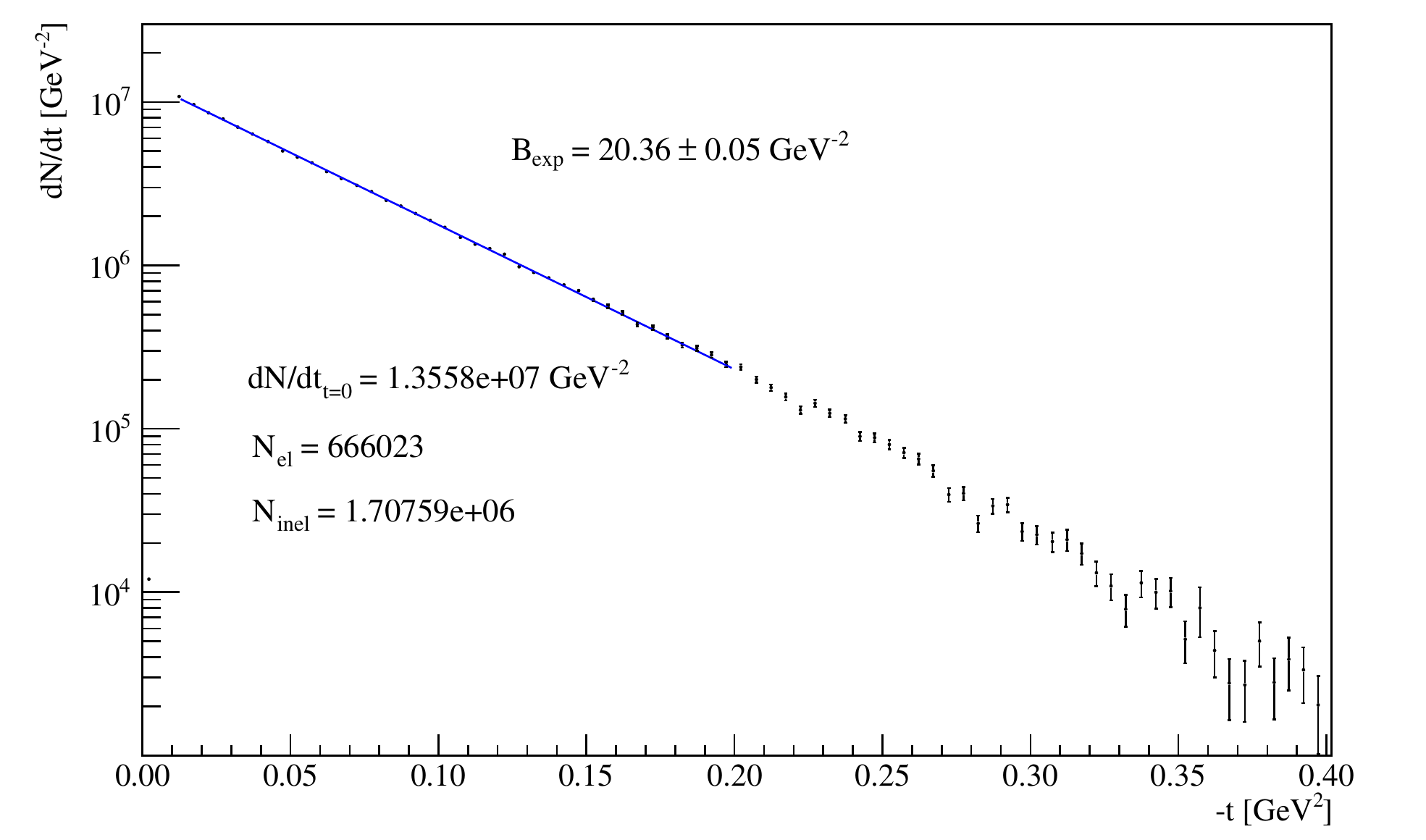}
		\includegraphics[trim = 0mm 0mm 0mm 0mm, clip,width=0.49\linewidth]{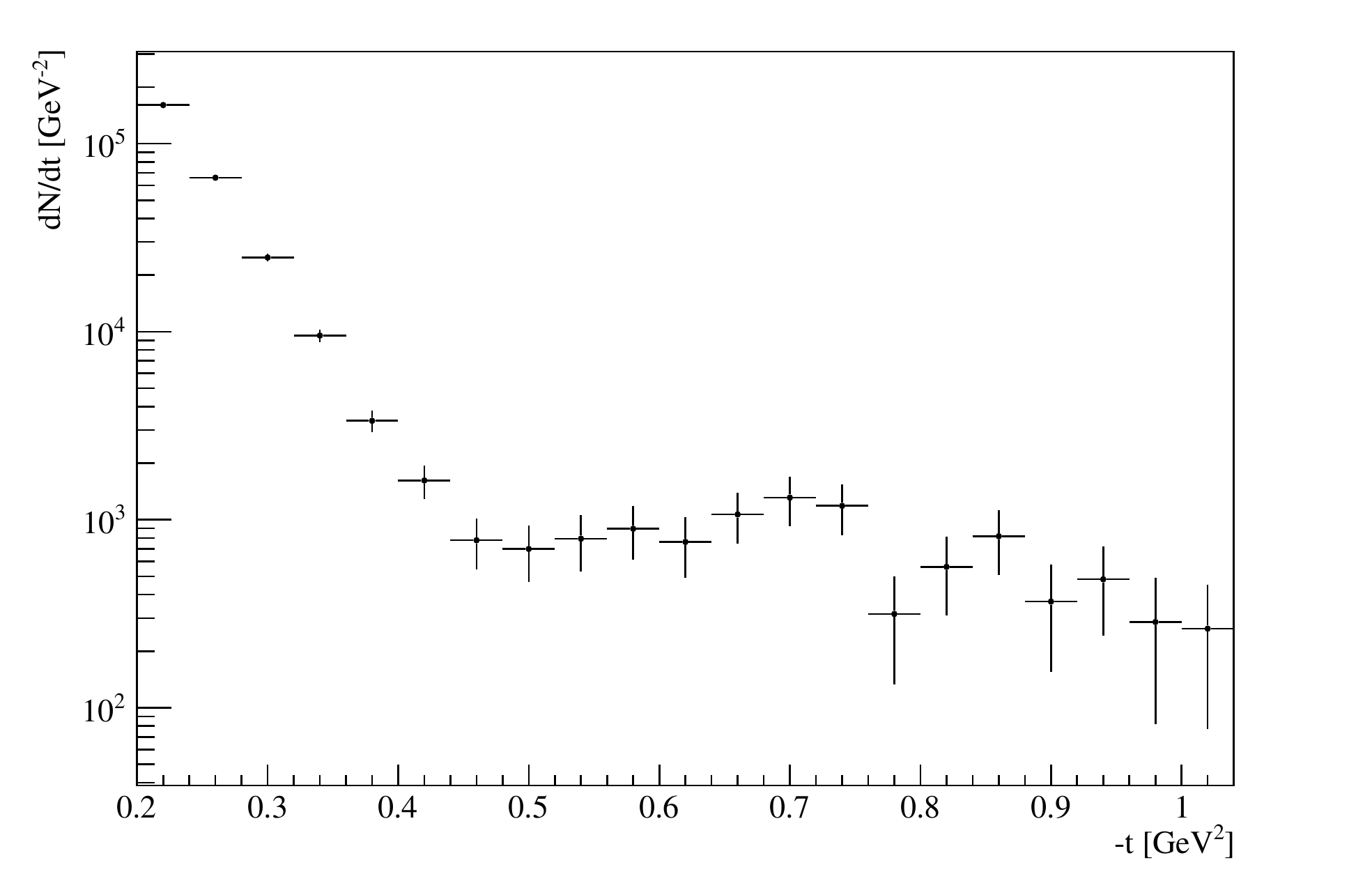}
		\caption{Differential elastic rate ${\rm d}N_{\rm el}/{\rm d}t$ at $\sqrt{s} = 13$~TeV (full physics corrections included) of dataset DS2 with the exponential fit
		between $|t|_{\rm min}$ and $|t|_{\rm max}$. The right panel shows the data in the dip region.}
		\label{differential_cross_section}
	\end{figure*}

	The stability of the fit has been verified by varying the lower $|t|$ bound. The observed systematic effect on the 
    slope and on the intercept at $t=0$ is negligible compared to the other systematic uncertainties listed in Table~\ref{analysis_corrections}. Assuming that the exponential parametrization holds also for $|t| < |t|_{\rm min}$, the value 
    $\left.{\rm d}N_{\rm el}/{\rm d}t\right|_{t=0}$ can be used to determine the total cross section using Eq.~(\ref{calculation_of_sigma_top}). The measurements performed at very high $\beta^{*}$ optics will allow the exploration of the $|t|$ region below 
      the present $|t|_{\rm min}$ to probe the Coulomb-nuclear interference or any other new effect. 

\subsection{The total cross section}
	
The measurements of the total inelastic rate $N_{\rm inel}$ and of the total nuclear elastic rate $N_{\rm el}$
(with its extrapolation to $t=0$, $\left.{\rm d}N_{\rm el}/{\rm d}t\right|_{t=0}$) are combined via the optical theorem to obtain
the total cross section in a luminosity independent way
	\begin{align}
		\sigma_{\rm tot}=\frac{16\pi(\hbar c)^{2}}{1+\rho^{2}}\cdot\frac{\left.{\rm d}N_{\rm el}/{\rm d}t\right|_{t=0}}{N_{\rm el} + N_{\rm inel}}\,,
		\label{calculation_of_sigma_top}
	\end{align}
where the parameter $\rho$ is the ratio of the real to the imaginary part of the forward nuclear elastic amplitude.

The total cross section measurements of the DS1 and DS2 data sets have been averaged according to their raw inelastic rate $N_{\rm inel,obs}$, which yields
	\begin{align}
	             \sigma_{\rm tot}=(110.6\pm3.4)~{\text{mb}}\,,
	\end{align}
when $\rho=0.1$ is assumed. The choice of $\rho=0.1$ in the present analysis is motivated by the results published in~\cite{Antchev:2298154}.

From the measured (and fully corrected) ratio of $N_{\rm el}$ to $N_{\rm inel}$ the luminosity- and $\rho$-independent ratios
	\begin{align}
	             \frac{\sigma_{\rm el}}{\sigma_{\rm inel}}=0.390\pm0.017,\,\,\,\frac{\sigma_{\rm el}}{\sigma_{\rm tot}}=0.281\pm0.009\,,
	\end{align}
The luminosity independent elastic and inelastic cross sections are derived by combining their ratio and sum
	\begin{align}
	             \sigma_{\rm el}=(31.0\pm1.7)~{\text{mb}},\,\,\,\sigma_{\rm inel}=(79.5\pm1.8)~{\text{mb}}\,.
	\end{align}
The measured physics quantities are also calculated for $\rho=0.14$ and the values are summarized in Table~\ref{physics_quantities}. 

	\begin{table*}[h]
		\centering
		\caption{The nuclear slope $B$, the cross sections and their systematic and statistical uncertainty. The physics quantities are the weighted average of the DS1 and
		DS2 measurements.}
            \begin{tabular}{| c  c  c  c |}
		\hline
                	Physics quantity			& \multicolumn{2}{c}{Value} 		&  Total uncertainty 	\\ 
                						& $\rho=0.14$	& $\rho=0.1$ 		&		 	\\ \hline
			$B$~[GeV$^{-2}$]			& \multicolumn{2}{c}{20.36}		& $5.3\cdot 10^{-2}\oplus$		0.18	= 0.19	\\
			$\sigma_{\rm tot}$~[mb]			& 109.5 & 110.6				& 					  3.4	\\ 
			$\sigma_{\rm el}$~[mb]			& 30.7  & 31.0				& 					  1.7	\\ 
			$\sigma_{\rm inel}$~[mb]		& 78.8  & 79.5				& 					  1.8	\\ 
			$\sigma_{\rm el}/\sigma_{\rm inel}$	& \multicolumn{2}{c}{0.390}		& 					  0.017	\\ 
			$\sigma_{\rm el}/\sigma_{\rm tot}$	& \multicolumn{2}{c}{0.281}		& 					  0.009	\\ \hline
            \end{tabular}
		\label{physics_quantities}
	\end{table*}

Figure~\ref{the_evolution_of_cross_sections} is the compilation of all the previous $\rm pp$ and $\rm p\bar{\rm p}$ total, elastic and inelastic measurements, together with a selected set of TOTEM
measurements. Figure~\ref{the_evolution_of_cross_sections_zoom}
shows a more detailed plot of the measurements in the range between 7 and 8 TeV with the TOTEM values for $\sigma_{\rm tot}$ obtained with different methods.

With the present measurement TOTEM has covered a range from $\sqrt{s}=2.76$~TeV to 13~TeV obtaining a variation of total cross-section from ($84.7\pm3.3$)~mb to ($110.6\pm3.4$)~mb~\cite{Antchev:2013paa,Antchev:2016vpy,Antchev:2011vs,Antchev:2013iaa,DIS2017_proceedings,Paper_2p76}.

The evolution of the elastic to total cross section ratio and the nuclear slope $B$ as function of $\sqrt{s}$ are shown in Figures~\ref{the_evolution_of_el_to_tot_ratio} and~\ref{the_evolution_of_B}.
The elastic to total cross section ratio increases with $\sqrt{s}$ is seen from Figure~\ref{the_evolution_of_el_to_tot_ratio}. In particular, the deviation at LHC energies of the nuclear slope from the
low energy linear extrapolation is clearly visible in Figure~\ref{the_evolution_of_B}.

	\begin{figure*}
			\includegraphics[width=0.95\linewidth]{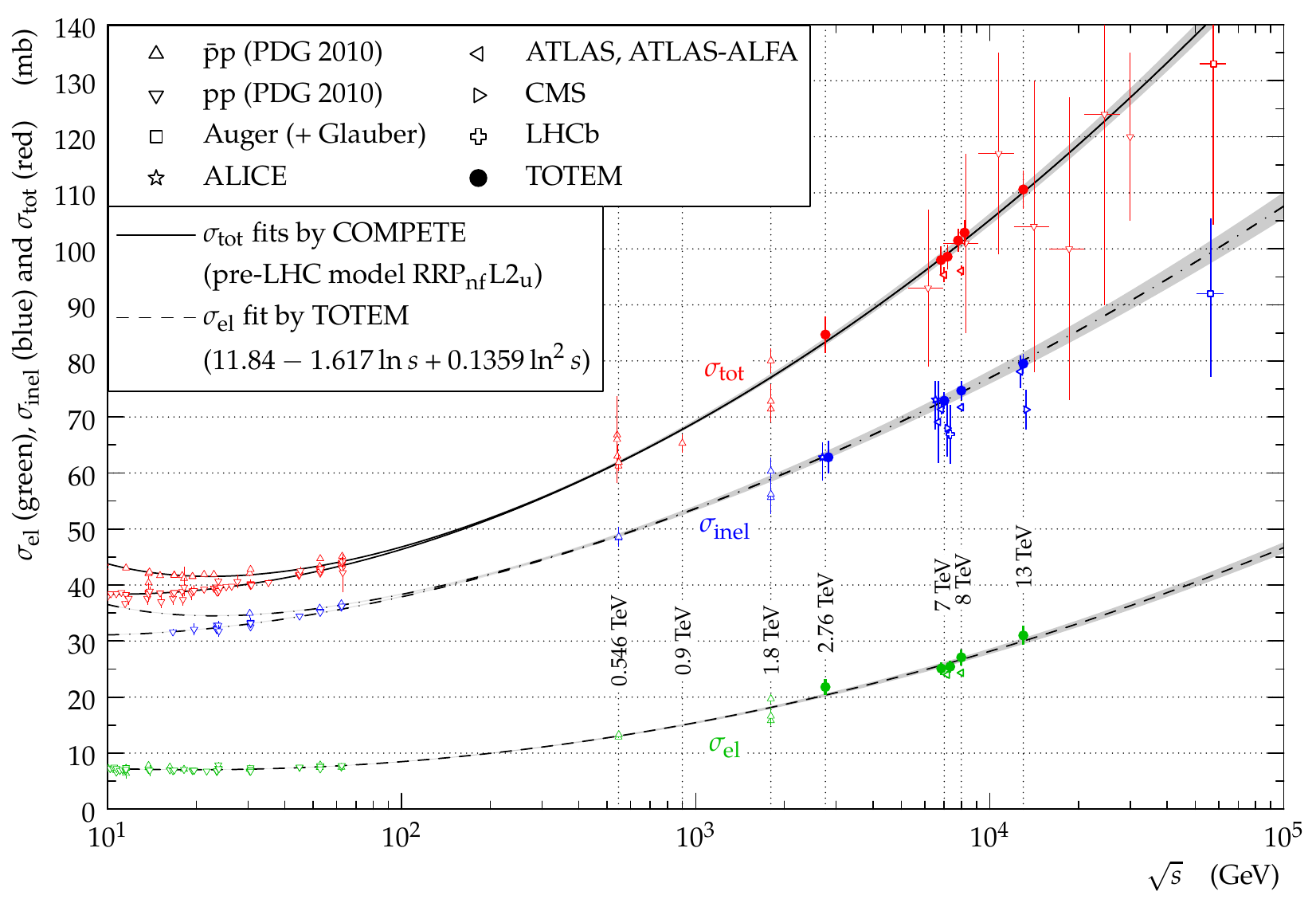}
			\caption{(color). Overview of elastic~ ($\sigma_{\rm el}$), inelastic~($\sigma_{\rm inel}$), total~($\sigma_{\rm tot}$) cross section for $\rm pp$ and $\rm p\bar{\rm p}$ collisions as a function of $\sqrt{s}$,
  			including TOTEM measurements over the whole energy range explored
			by the LHC~\cite{Antchev:2013paa,DIS2017_proceedings,Paper_2p76,Antchev:2015zza,Antchev:2016vpy,Antchev:2013iaa,Nakamura:2010zzi,Abelev:2012sea,atlas_report_1,cms_report_1,Collaboration:2012wt,Aad:2014dca,Aaboud:2016ijx,Aaij:2014vfa,Aaboud:2016mmw,CMS:2016ael,Antchev:2013gaa}. Uncertainty band on theoretical models and/or fits are as described in the
			legend. The continuous black lines (lower for $\rm pp$, upper for $\rm p\bar{\rm p}$) represent the best
				fits of the total cross section data by the COMPETE collaboration~\cite{Cudell:2002xe}. The dashed line results from a fit of the elastic scattering data.
				The dash-dotted lines refer to the inelastic cross section and are obtained as the difference between the continuous and dashed fits.}
			\label{the_evolution_of_cross_sections}\vspace{2mm}
	\end{figure*}

	\begin{figure*}
			\centering
			\includegraphics[width=0.8\linewidth]{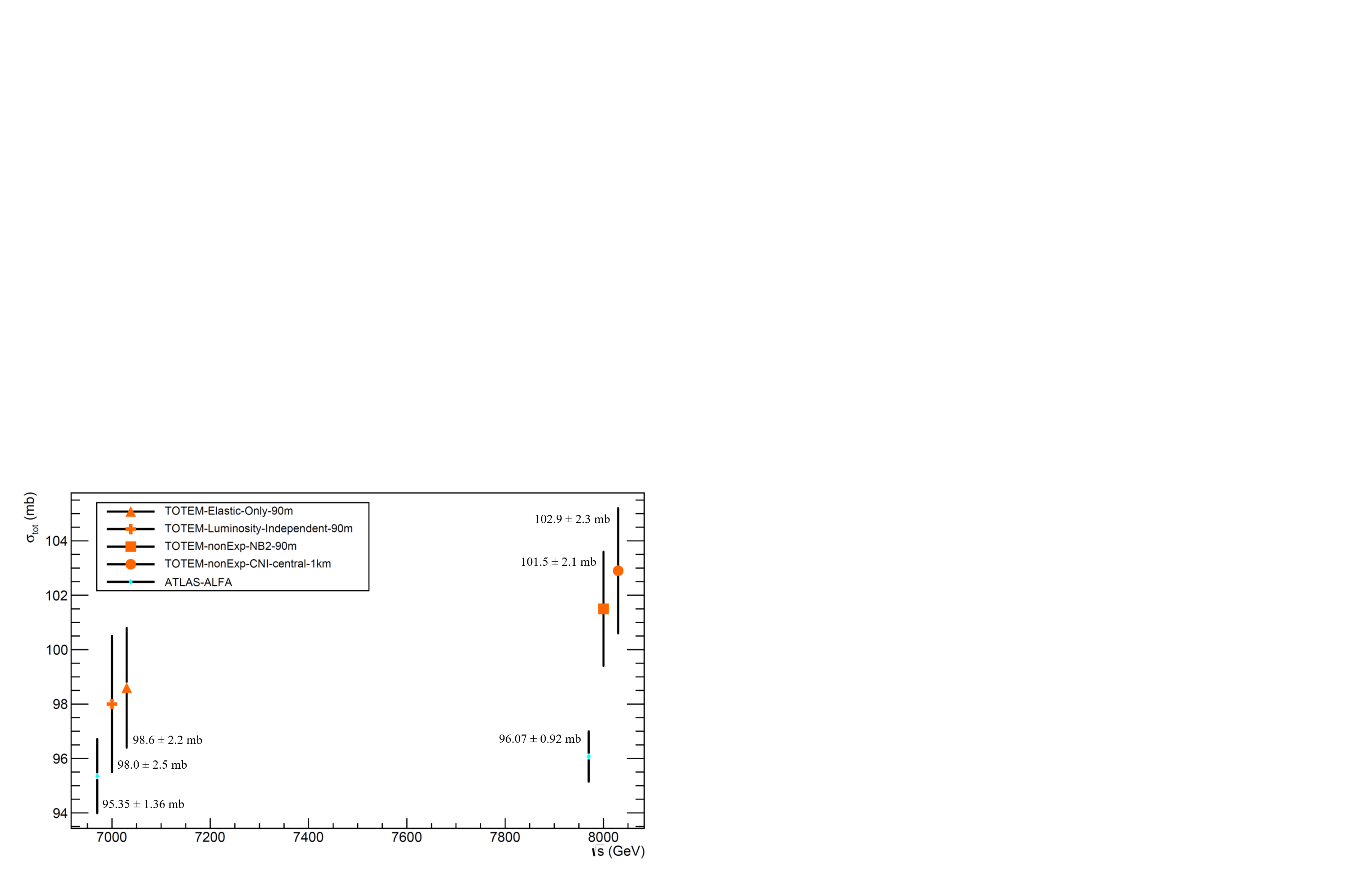}
			\caption{(color). Focus on the 7~-~8 TeV range showing the comparison of the pairs of TOTEM measurements which represent the broadest
				 exploration of different methods, data sets, $t$-range (with or without Coulomb-nuclear interference) and
 				descriptions of the nuclear slope with the ATLAS-ALFA measurements
				~\cite{Antchev:2013iaa,Antchev:2015zza,Antchev:2016vpy,Aaboud:2016ijx,Antchev:2013gaa,Aad:2014dca}.}
			\label{the_evolution_of_cross_sections_zoom}\vspace{2mm}
	\end{figure*}

	\begin{figure*}
			\includegraphics[width=0.95\linewidth]{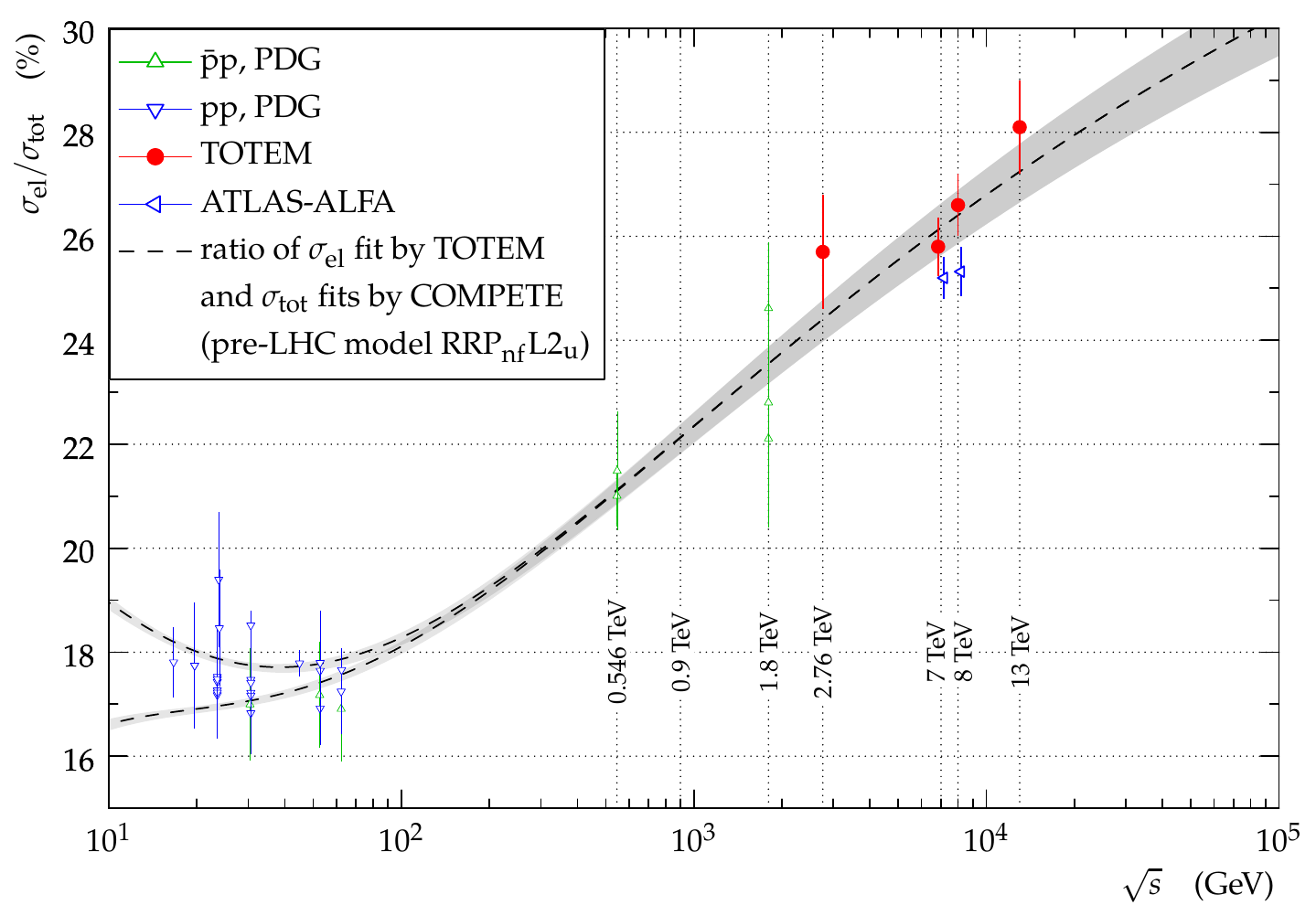}
			\caption{(color). The elastic to total cross section ratio for $\rm pp$ and $\rm p\bar{\rm p}$ collisions as a
			function of $\sqrt{s}$~\cite{Aad:2014dca,Aaboud:2016ijx,Nakamura:2010zzi,DIS2017_proceedings,Paper_2p76,Antchev:2013iaa,Antchev:2013paa}.}
			\label{the_evolution_of_el_to_tot_ratio}\vspace{2mm}
	\end{figure*}

	\begin{figure*}
			\includegraphics[width=0.95\linewidth]{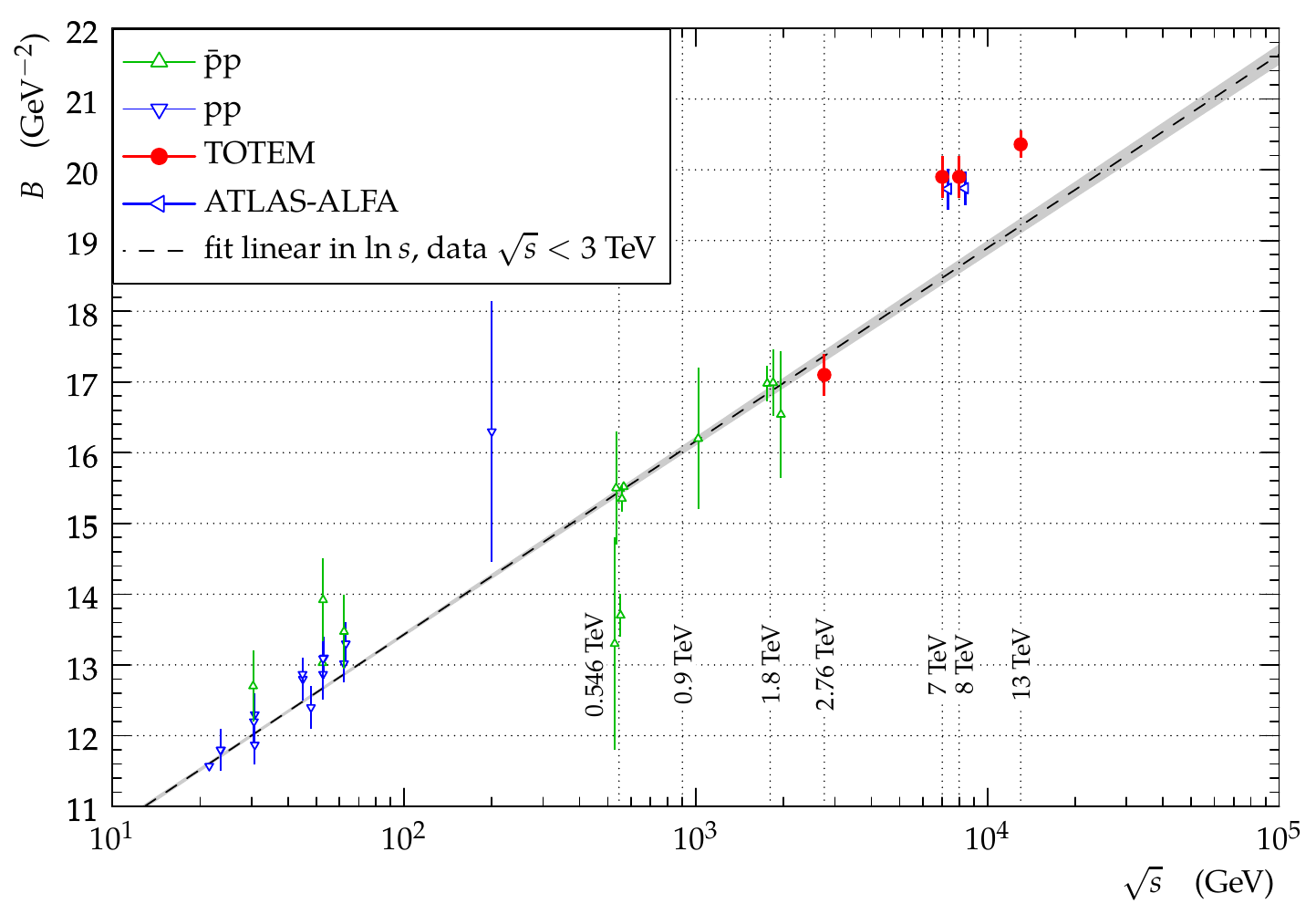}
			\caption{(color). The nuclear slope $B$ for $\rm pp$ and $\rm p\bar{\rm p}$ elastic scattering as a function of $\sqrt{s}$.
 			It should be understood that while $B$ is defined at $t=0$, the experimental measurements are actually averaging the slope, 
 			hence they depend on the chosen $t$-range and on the deviations of the data from a pure exponential.
 			While fluctuations beyond the experimental error bars should thus be expected, the deviation for $\sqrt{s}>3$~TeV from the linear extrapolation is highly significant~\cite{Aad:2014dca,Aaboud:2016ijx,Nakamura:2010zzi,DIS2017_proceedings,Paper_2p76,Antchev:2013gaa,Antchev:2013paa}.}
			\label{the_evolution_of_B}\vspace{2mm}
	\end{figure*}

\section*{Acknowledgments}

We are grateful to the beam optics development team
for the design and the successful commissioning of the
high $\beta^{*}$ optics and to the LHC machine coordinators for
scheduling the dedicated fills.

This work was supported by the institutions listed on the
front page and partially also by NSF (US), the Magnus Ehrnrooth Foundation (Finland), the Waldemar von
Frenckell Foundation (Finland), the Academy of Finland,
the Finnish Academy of Science and Letters (The Vilho 
Yrj\"o and Kalle V\"ais\"al\"a Fund), the OTKA NK 101438 and the EFOP-3.6.1-16-2016-00001 grants
(Hungary). Individuals have received support from Nylands nation vid Helsingfors universitet (Finland),
MSMT CR (the Czech Republic), the J\'{a}nos Bolyai Research Scholarship of
the Hungarian Academy of Sciences and the NKP-17-4 New National Excellence Program of the
Hungarian Ministry of Human Capacities.

\clearpage
\bibliographystyle{utphys}
\bibliography{mybib}

\end{document}